# Intrinsic Interstellar Absorption in two Radio Galaxies at z=0.8

Sperello di Serego Alighieri[1], Andrea Cimatti[1] and Sandra Savaglio[2]

[1] Osservatorio Astrofisico di Arcetri, Firenze, Italy
[2] Università della Calabria, Cosenza, Italy

**Abstract.** We have found a strong absorption feature at 2598Å in the spectra of two distant radio galaxies, which we interpret as due to interstellar FeII. We discuss this interpretation and the possibility of obtaining important information on the evolution of the ISM in galaxies, to be compared with the results of the study of intervening and associated absorptions in quasars.

## 1   Introduction

Being the most distant stellar systems known (Lacy et al. 1994), radio galaxies can provide essential information for the study of the formation and early evolution of galaxies, both for their stellar and interstellar content. So far information on the ISM in distant galaxies has come from bright emission lines by ionized gas, from infrared emission presumably by dust, from a few CO detections from molecular clouds, from the polarization produced by dust and electron scattering, and from the so–called intervening absorptions in quasar spectra. These informations are complementary, in the sense that they sample different phases of the ISM at different locations in the galaxy.

A direct detection of intrinsic interstellar absorption in a distant galaxy has not been obtained yet, because of the very faint continuum of these objects. However such a detection would be important to study the evolution of the ISM — its metallicity and depletion in particular — and to compare the results with those obtained from intervening absorptions in quasar spectra. Radio galaxies are usefull in this respect, both because they can be very distant and because they have a relatively higher continuum level in the ultraviolet, where interstellar lines are stronger. Furthermore, if one is prepared to accept that radio galaxies are just quasars viewed from a different angle as foreseen by the AGN Unified Model (Antonucci, 1993), intrinsic interstellar absorptions in radio galaxies can be compared directly with the so–called associated absorptions in quasar spectra, giving information on the ISM around the most luminous AGN.

We discuss here our recent detection (di Serego Alighieri et al. 1994) of a strong ultraviolet absorption in the spectra of two distant radio galaxies.

## 2  Interstellar absorption at z=0.8?

During our spectropolarimetric study of distant radio galaxies we have detected a strong absorption feature at 2598Å in the spectra of the radio galaxies 3C 226 (z=0.818) and 3C 277.2 (z=0.766). Because of the needs of spectropolarimetry we have two long exposure spectra, of 45 minutes each, for both galaxies. The absorption is detected in each of the 4 spectra. Since the spectra are all very similar, we show in Figure 1 a section of the summed spectrum in rest-frame wavelengths. The absorption feature has a large rest-frame equivalent width ($8 \pm 2$Å).

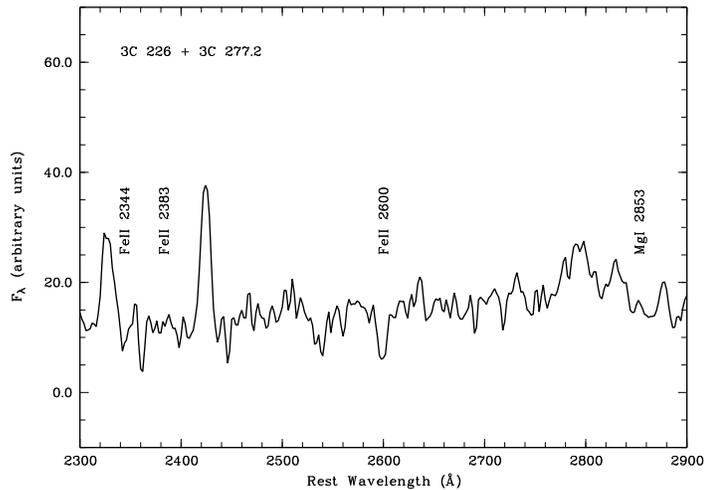

**Fig. 1.** Summed spectra of 3C 226 and 3C 277.2. The positions of strong ultraviolet interstellar absorptions are marked.

The line is detected at the same rest-frame wavelength in the spectra of the two galaxies, which are at a different redshift and are separated by a very large angle in the sky. Therefore we believe that it is very unlikely that the line is due to an intervening absorber and we conclude that it should be produced by intrinsic absorption in the galaxies.

Could the absorption be of stellar origin? The only stellar absorption at the right wavelength is FeII2600 which is present in F and G type stars (Fanelli et al. 1990). Nevertheless these stars emit a very small fraction of their radiation in the ultraviolet and could hardly explain the ultraviolet to visual colour of the galaxies. Moreover we have evidence from the spectropo-larimetry that the dilution of the polarized flux is small in the ultraviolet around the MgII2800 emission line (di Serego Alighieri et al. 1994). Therefore a stellar absorption like FeII2600 in F and G stars could not have such a high equivalent width in the total spectrum, and we exclude that the absorption is of stellar origin.

We tentatively attribute the absorption feature to interstellar FeII in the two galaxies. In support of this interpretation we bring the fact that indeed FeII has a multiplet around 2600Å, whose strongest components, at 2600.2 and 2586.5Å, we cannot resolve, since the resolution of our spectra is about 15Å. FeII2600.2 is the strongest ultraviolet interstellar absorption after the lines of MgII2797+2804 (Kinney et al. 1993), which we cannot detect in our spectra since they are filled by emission.

However there are two problems with the interstellar FeII interpretation. First, in our spectra we do not detect so clearly other strong interstellar ultraviolet lines of FeII and other ions, although we cannot completely exclude their presence, given the S/N ratio of our spectra (see Fig. 1). Second, the rest–frame equivalent width of the absorption feature is much larger than what is normally observed for FeII2586+2600 in the ISM of our galaxy ($\sim$2Å, Kinney et al. 1993). This fact could be due to the different conditions of the ISM in distant radio galaxies, such as a larger column density, different ionization conditions and a lower depletion. Indeed a very strong absorption at 2600Å is present in the IUE spectrum of the supernova SN 1986G, which exploded in the dust lane of the powerful radio galaxy Cen A. Also several components at different velocities, which we see blended in our low resolution spectra, could increase the equivalent width above galactic levels.

The possibility of a lower depletion than the galactic one can be associated with the unification of quasars and radio galaxies. Our polarimetric studies have demonstrated that a large fraction of the ultraviolet continuum from distant radio galaxies is actually scattered and is probably radiation which comes originally from a quasar in the nucleus, hidden from direct view. Therefore this ultraviolet radiation, before and after the scattering, travels through considerable distances inside the strong radiation cone of the quasar, a region where dust can hardly survive, thereby explaining a low depletion.

If radio galaxies are indeed quasars viewed from outside the radiation cones, it would be interesting to compare our results with those on associated absorptions in quasars. The latter seem to have a much higher ionization level: for example CIV is seen, but not FeII (e.g. Savaglio et al. 1994). This could be attributed to the fact that our line of sight to radio galaxies travels through their ISM also outside the radiation cone, where ionization is lower.

## 3   Outlook

We certainly need to confirm our tentative detection of interstellar absorption in distant radio galaxies. If this is confirmed with spectra taken specially for this purpose, then a further study of these absorptions, in particular with the new 8-10m class telescopes, would allow to obtain important information on the ISM at early cosmological epochs and on its evolution. In particular we could obtain constraints on column densities and metallicity, on the presence of various components at different velocities, on the ultraviolet photoionizing source from the relative abundance of the different ions of the same element, and on the depletion.

We have shown that it is becoming possible to study interstellar absorptions in distant galaxies directly, allowing the study of an essential component of the formation and evolution of galaxies and a useful comparison with the results from intervening and associated absorptions in quasars.

*Acknowledgements.* We thank Sofia Randich and Roberto Viotti for usefull discussions on the properties of interstellar absorption lines.